\documentclass[prl,superscriptaddress,floatfix,nobalancelastpage,twocolumn]{revtex4}
\usepackage{bm}
\usepackage{amsmath}
\usepackage{verbatim}
\usepackage{bm,braket}
\usepackage{color}
\usepackage{float}
\usepackage{graphicx}
\usepackage{multirow}
\usepackage{array}
\usepackage{amssymb}
\usepackage{sidecap}
\usepackage{enumitem}
\usepackage[toc,page]{appendix}
\usepackage{epstopdf}
\usepackage[caption=false]{subfig}
\usepackage{sidecap}
\usepackage{enumitem}
\usepackage[toc,page]{appendix}
\usepackage{epstopdf}
\usepackage{subfig}
\usepackage{sidecap}
\usepackage[utf8]{inputenc}
\usepackage[compact]{titlesec}
\usepackage{tikz}

\makeatletter
\def\@seccntformat#1{%
        \expandafter\ifx\csname c@#1\endcsname\c@section\else
        \csname the#1\endcsname\quad
        \fi}
\makeatother

\definecolor{myBlue}{rgb}{0.0430,0.5156,0.7773}
\definecolor{light-gray}{gray}{0.95}

\begin{document}

        \frenchspacing 
        \abovedisplayskip=8pt
        \abovedisplayshortskip=8pt
        \belowdisplayskip=8pt
        \belowdisplayshortskip=8pt
        \arraycolsep=100pt
        
        \title{Hyperbolic Metamaterial Nano-Resonators Make Poor Single Photon Sources}

        \author{Simon Axelrod}
        \affiliation{Department of Physics, Engineering Physics and Astronomy,
                Queen's University, Kingston, Canada, K7L 3N6}
                
                   \author{Mohsen Kamandar Dezfouli}
        \affiliation{Department of Physics, Engineering Physics and Astronomy,
                Queen's University, Kingston, Canada, K7L 3N6}
                
                 \author{Herman M. K. Wong}
               \affiliation{Photonics Group, Edward S. Rogers Sr. Department of Electrical and Computer Engineering, University of Toronto, Toronto, Canada, M5S 3H6}

\author{Amr S. Helmy}
 \affiliation{Photonics Group, Edward S. Rogers Sr. Department of Electrical and Computer Engineering, University of Toronto, Toronto, Canada, M5S 3H6}

 \author{Stephen Hughes}        
 \email{shughes@queensu.ca}
        \affiliation{Department of Physics, Engineering Physics and Astronomy,
                Queen's University, Kingston, Canada, K7L 3N6}
 
        \begin{abstract}
We study the optical properties of quantum dipole emitters coupled to hyperbolic metamaterial nano-resonators using a semi-analytical quasinormal mode approach. We show that coupling to metamaterial nano-resonators can lead to significant Purcell enhancements that are nearly an order of magnitude larger than those of plasmonic resonators with comparable geometry. However, the associated single photon output $\beta$-factors  are extremely low (around 10\%), far smaller than those of comparable sized metallic resonators (70\%). Using a quasinormal mode expansion of the photon Green function, we describe how\ the low $\beta$-factors are due to increased Ohmic quenching arising from redshifted resonances, larger quality factors, and stronger confinement of light within the metal.
In contrast to current wisdom, these results
suggest that  hyperbolic metamaterial nano-structures make poor choices for single photon sources.
        \end{abstract}
        
\maketitle

\begin{figure}[t]
\begin{tikzpicture}
\node[anchor=south east,inner sep=0] at (4,0.2){\includegraphics[height=0.4\columnwidth,trim={0cm 0cm 0cm 0cm},clip]{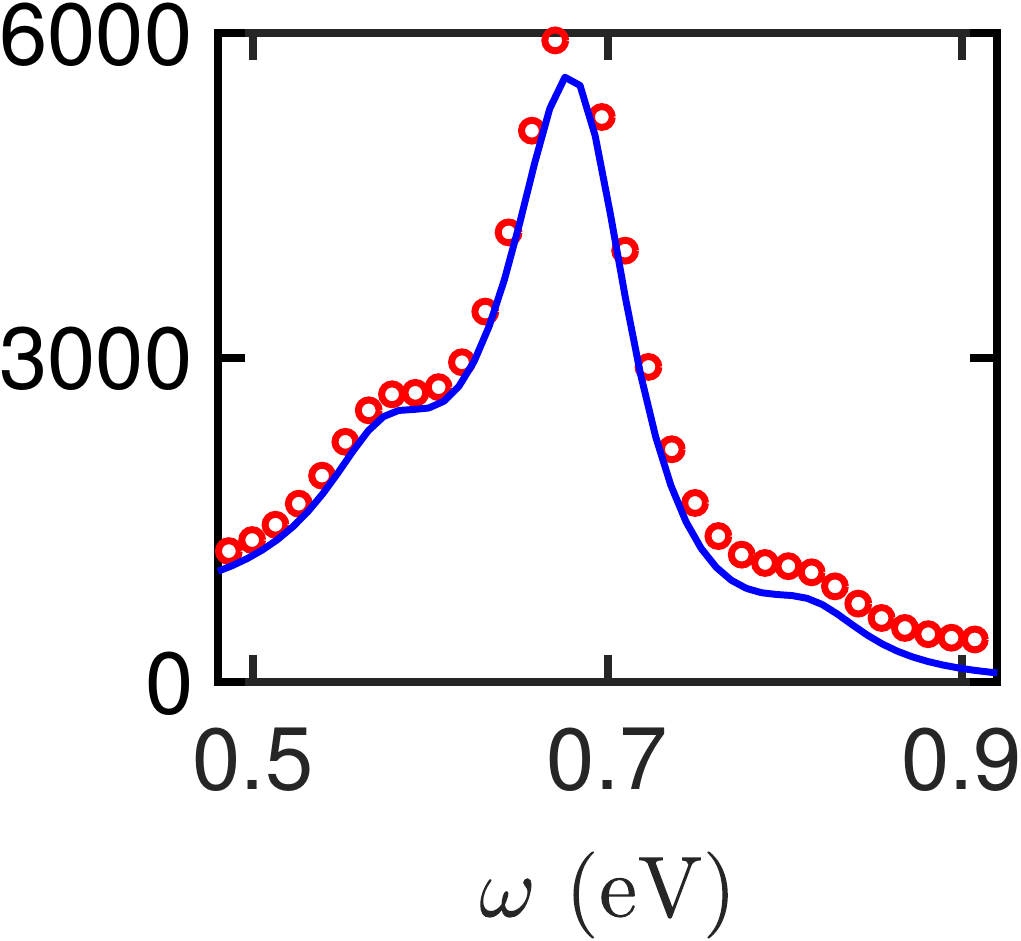}};

\node[anchor=south west,inner sep=0] at (-4,0.2){\includegraphics[height=0.4\columnwidth,trim={0cm 0cm 0cm 0cm},clip]{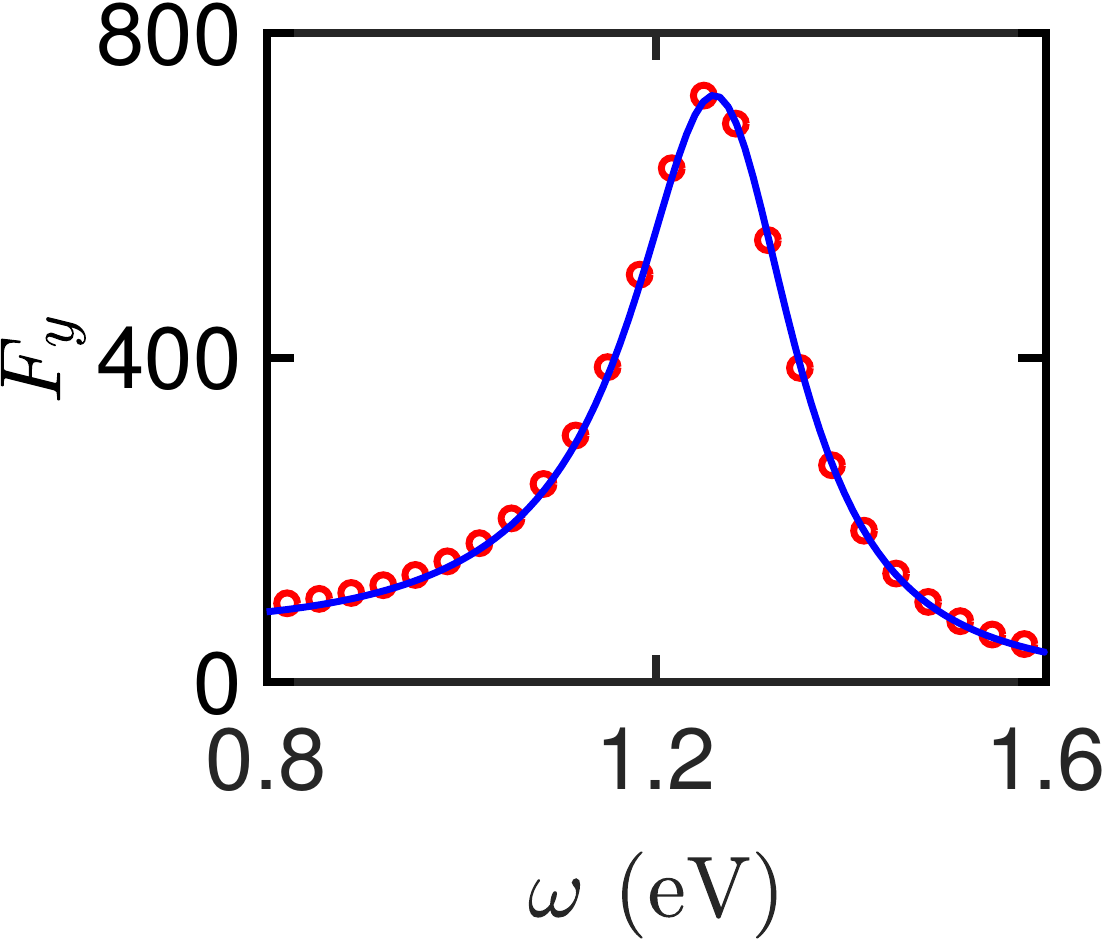}};
\node[anchor=north,inner sep=0] at (0.2,5.5){\includegraphics[width=0.86\columnwidth,trim={0cm 0cm 0cm 0cm},clip]{pf_plots_half}};
\node at (-2.5,5.9) {(a)};
\node at (1.6,5.9) {(b)};
\node at (-2.5,3.2) {(c)};
\node at (1.6,3.2) {(d)};
\begin{scope}[shift={(-7.7,1)},color=black]
\draw[line width=0.8,-stealth] (6,4.5) -- (6.24,4.4856);
\node at (5.82,4.2) {$x$};
\draw[line width=0.8,-stealth] (6,4.5) -- (5.892,4.32);
\node at (6.38,4.42) {$y$};
\draw[line width=0.8,-stealth] (6,4.5) -- (6,4.7688);
\node at (5.98,4.92) {$z$};
\end{scope}
\end{tikzpicture}
                \vspace{-0.0cm}
        \caption{(a) Schematic of a gold nano-dimer resonator. A $y$-polarized quantum dipole is shown in the gap centre. (b) Schematic of an HMM dimer with 7 layers of gold and 6 layers of dielectric (blue). (c) Purcell factor for a $y$-polarized dipole in the gap centre of a gold nano-dimer, obtained with full dipole calculations (red circles) and a QNM expansion (solid blue). (d) Purcell factor as in (c), but for an HMM dimer of metal filling fraction $f_m = 0.2$.}
        \label{fig:schematic}
        \centering
\end{figure}

\begin{figure}[t]
        \centering\includegraphics[width=1.0\columnwidth,clip=true,trim= 0cm 0cm 0cm 0cm]{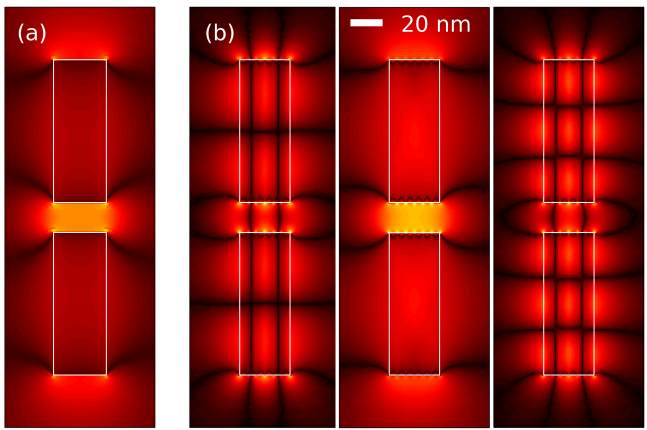}
\vspace{-0.4cm}
        \caption{(a) QNM field profile $\vert \tilde{\mathbf{f}}_y(0,y,z) \vert $ for the dominant mode of a plasmonic dimer. The edges of the dimer are shown in white. (b) QNM profile for the three dominant modes of an HMM dimer with filling fraction $f_m = 0.2$, with eigenfrequencies increasing from left to right. Brighter colours indicate stronger fields.} 
        \label{fig:qnm_profile}
\end{figure}

\textit{Introduction.} 
Engineered cavity structures allow for tight confinement of light and the enhancement of its interaction with matter. In particular, solid state structures such as photonic crystals \cite{photonic_crystal_1,photonic_crystal_2}, slow-light waveguides \cite{slow_light_waveguide}, plasmonic nano-structures \cite{SE_plasmon_1,SE_plasmon_2_equals_gamman_nr,SE_plasmon_3} and metamaterial resonators \cite{SE_HMM_1,SE_HMM_2} allow for the enhancement of the photon local density of states (LDOS) of embedded quantum emitters, thereby increasing their spontaneous emission rates via the Purcell effect \cite{purcell_original}. Such enhancement finds application in areas such as molecule sensing \cite{molecule_sensing}, high-resolution imaging \cite{imaging1,imaging2}, energy harvesting \cite{energy_harvesting_1,energy_harvesting_2}, nonlinear optics \cite{nonlinearoptics}, and single photons  \cite{single_photon}. \\ 
\indent A new class of optical materials known as hyperbolic metamaterials (HMMs) offers the possibility of achieving extreme confinement of light and increased interaction with matter over a broad spectral range \cite{seminal_smith_1,seminal_smith_2, jacob_review}. Such materials consist of both metal and dielectric parts, and are typically described as having an anisotropic dielectric tensor within an effective medium description. The dielectric tensor elements $\varepsilon_{\parallel}$ and $\varepsilon_{\perp}$ (parallel and perpendicular to the axis of anisotropy, respectively) are of opposite sign, corresponding to metallic or dielectric properties along different axes. For an HMM that is anisotropic along the $z$-axis, for example, the electromagnetic dispersion relation is given by %
\begin{align}
 \frac{k_x^2 + k_y^2}{\varepsilon_{\parallel}} + \frac{k_z^2}{\varepsilon_{\perp}} = \frac{\omega^2}{c^2},
\end{align} 
where $\mathbf{k}$ is the wavevector, $\omega$ is the angular frequency, and $c$ is the speed of light. Since   $\varepsilon_{\parallel}$ and $\varepsilon_{\perp}$ are of opposite sign, surfaces of constant frequency are hyperbolic, extending to very large values of $k$. The resulting momentum mismatch between HMM and free-space electromagnetic fields results in strong confinement of light around the structure \cite{zhang_experimental}. Moreover, the isofrequency dispersion implies that dipole emitters can couple to a large range of $k$-states at a single frequency, thereby increasing the number of possible decay paths and thus the spontaneous emission rate \cite{jacob_review}. Metamaterial waveguides have also been shown to provide enhanced Purcell factors and Lamb shifts through the associated slow light modes \cite{slow_light_waveguide}. \\
\indent Many applications in HMM and plasmonic nanophotonics require Purcell enhancements that are radiative in nature \cite{jacob_review, jacob2010, noginov, sipe2011, Ni, jacob2012, experimental_HMM}, and it is often of fundamental importance to minimize non-radiative metallic losses. The minimization of such losses is one of the biggest unresolved issues in plasmonics and metamaterial science, limiting nearly every potential application in these fields \cite{khurgin}. While several works have sought to mitigate such losses \cite{khurgin,loss_and_gain,low_loss_plasmonics,graphene_plasmonics}, the issue remains an outstanding concern. Despite the importance of analyzing loss in plasmonic and metamaterial resonators, there has been little conclusive analysis of the latter. Theoretical studies have argued that the Purcell enhancement in simple HMM slabs is radiative in nature \cite{jacob_review,noginov,jacob2012}, and experimental work \cite{experimental_HMM} has compared radiative and non-radiative decay in metal and HMM slab structures, but a thorough investigation of quenching in HMMs has not been performed. The role of Ohmic damping has been compared in HMM and metal cavities \cite{Zhang}, but energy loss has not. The superior ability of HMMs to engineer radiative decay has also been questioned theoretically \cite{2014_prl,moscow}. An analytical description of radiative and non-radiative decay in HMM and metal resonators is thus of great interest. \\
\indent In this Letter, we study metal and HMM nano-resonators for application in single photon emission, providing a representative analysis of non-radiative loss in such structures. We compare the associated spontaneous emission enhancements and single photon output $\beta$-factors (the probability of emitting a photon via radiative decay) using a semi-analytical Green function (GF) approach. We first show that the GF of a complex, multi-layered HMM resonator can be simply and accurately described in terms of its quasinormal modes (QNMs), the optical modes for an open dissipative cavity structure \cite{QNM_cavity,ACS_photonics}. We report greatly enhanced spontaneous emission rates in HMMs (up to an order of magnitude greater than those of metal resonators with comparable geometry), but surprisingly, significantly lower $\beta$-factors. Using a QNM approach, we show that this increased quenching is due to a combination of redshifted resonances, larger quality factors, and stronger confinement of light within the metal regions of HMMs. We conclude that HMM resonators are characterized by greatly enhanced Purcell factors that are always accompanied by smaller $\beta$-factors, making them poor choices for single photon sources and radiative decay engineering.\\
\indent \textit{QNM Green Function Expansion.} The light-matter interactions are  rigorously described in terms of the  photon GF. For example,  the LDOS enhancement $\rho (\mathbf{r}_a,\omega)/ \rho^{h}(\mathbf{r}_a,\omega)$ of an $\mathbf{n}_a$-polarized emitter at position $\mathbf{r}_a$ is given by the ratio $\mathrm{Im} \{ \mathbf{n}_a {\cdot} \mathbf{G}(\mathbf{r}_a,\mathbf{r}_a;\omega) {\cdot} \mathbf{n}_a \} /\, \mathrm{Im} \{ \mathbf{n}_a {\cdot} \mathbf{G}^h(\mathbf{r}_a,\mathbf{r}_a;\omega) {\cdot} \mathbf{n}_a) \}$~\cite{novotny}, where $\mathbf{G}$ is the GF and $h$ denotes a homogeneous background medium. Within the weak coupling regime, the LDOS enhancement represents the Purcell factor. Moreover, the GF can be used to quantify the non-radiative decay rate, through \cite{two_dots,SE_plasmon_2_equals_gamman_nr}
\begin{align}
\label{eq:gamma_nr}
\gamma^{\rm{nr}}(\mathbf{r}_{a},\omega) = \frac{2}{\hbar \omega \varepsilon_0} \int_{V} {\rm Re} \{ \mathbf{j(r)} {\cdot} {\mathbf{G}^*({\bf r},{\bf r}_a;\omega)} {\cdot} {\bf d}_a \}\, \mathrm{d}\mathbf{r},
\end{align}
where  $\mathbf{d}_a = d \mathbf{n}_a$ is transition dipole of the emitter, and $\mathbf{j(r)} =  \ \omega \kern 0.1667em {\rm Im} \{ \mathbf{\varepsilon}({\bf r},\omega) \} \mathbf{G}({\bf r},{\bf r}_a;\omega) {\cdot} {\bf d}_a$ is the induced current density within the scattering geometry. \\
\indent The GF is known analytically in a few simple cases, but in general must be obtained numerically. Full numerical solutions of Maxwell's equations can be obtained for a radiating dipole emitter located at position $\mathbf{r}_a$ in a given photonic environment. Using the electric field solution at general positions $\mathbf{r}$, one can obtain the two space-point GF $\mathbf{G(r,r}_a;\omega)$, \cite{dipole_sim_1,SE_plasmon_1,dipole_sim_3}, and therefore the LDOS at the dipole location ($\propto \mathrm{Im} \{\mathbf{G}(\mathbf{r}_a,\mathbf{r}_a;\omega) \}$). Note that one can also obtain the single photon output $\beta$-factor by calculating the proportion of the total dipole power that is radiated in the far field. However, the dipole approach requires another lengthy simulation to quantify the relevant physics at each new position. Instead, the GF may be expanded in terms of the QNMs of the scattering geometry. The QNMs $\tilde{\mathbf{f}}_{\mu}$ are the source-free solution to Maxwell's equations with open boundary conditions \cite{QNM_def_1,QNM_def_2}, with a discrete set of complex eigenvalues $\tilde{\omega}_{\mu} = \omega_{\mu} - i \gamma_{\mu}$, and associated quality factors $Q = \omega_{\mu} / 2 \gamma_{\mu}$. Due to the outgoing boundary conditions, QNMs diverge (exponentially) in space \cite{QNM_def_1,QNMs_diverge}, but their norm is still finite, and can be obtained in a number of equivalent ways (e.g. \cite{lai,sauvan,NORM,muljarov})
Within the resonator of interest \cite{QNM_def_1}, the transverse part of the GF can be written as an expansion of its QNMs, through \cite{QNM_def_2} $\mathbf{G}^{\mathrm{T}}\mathbf{(r,r';\omega)} = \sum_{\mu} (\omega^2 / 2 \tilde{\omega}_{\mu}(\tilde{\omega}_{\mu} - \omega)) \tilde{\mathbf{f}}_{\mu}(\mathbf{r}) \tilde{\mathbf{f}}_{\mu}(\mathbf{r'})$. For positions near metallic resonators (but outside the regime of quasi-static quenching), the GF can be accurately approximated by the same expansion \cite{NJP}, with the sum greatly reduced to the contribution of one or a few dominant modes near the main cavity resonance \cite{ACS_photonics}. Thus obtaining the dominant QNMs is usually sufficient for obtaining the GF as a function of frequency and position around the resonator. The GF and QNMs can then be used in various quantum optics formalisms \cite{photonic_crystal_2,cole_mnp,two_dots}, providing the starting point for an analytical and rigorous description of light-matter interactions.

\textit{HMM Nano-Dimers.} For practical purposes we analyze a parallelepiped nano-scale HMM dimer with 7 layers of gold and 6 layers of dielectric, anisotropic along the $z$-axis (Fig. \ref{fig:schematic}b), but our general findings below apply to all HMM geometries that we have tried (see below). The dimer configuration enhances Purcell factors in the gap through the bonding effect, and minimizes non-radiative quenching by drawing fields out of the metal \cite{ol}. The length of each parallelepiped is $95$ nm ($y$-axis), and the width and depth are $35$ nm ($x$- and $z$-axes). We set the gap size to $20$ nm in order to maximize the Purcell factor while minimizing non-radiative quenching. We set $\varepsilon = 2.9$ for the dielectric (similar to MgO) and $\varepsilon^h = 2.25$, and  use a Drude model for gold, $\varepsilon(\omega) = 1 - \omega_p^2 / (\omega (\omega+i \gamma))$. We set the plasmon frequency $\omega_p = 1.202 \times 10^{16} \ \mathrm{rad}/\mathrm{s}$ and collision rate $\gamma = 1.245 \times 10^{14} \ \mathrm{rad}/\mathrm{s}$, with parameters obtained by fitting experimental data for thin film gold in the frequency regime of interest \cite{experimental_data}. The use of a classical permittivity has been shown to be valid for material layers as thin as 1 nm \cite{nature_drude,drude2,drude3,drude4,drude5}.\\ 
\begin{figure}[t]
        \begin{tikzpicture}[      
        every node/.style={anchor=south west,inner sep=0pt},
        x=1mm, y=1mm,
        ]   
        \node (fig1) at (0,0)
        {\includegraphics[width=\columnwidth]{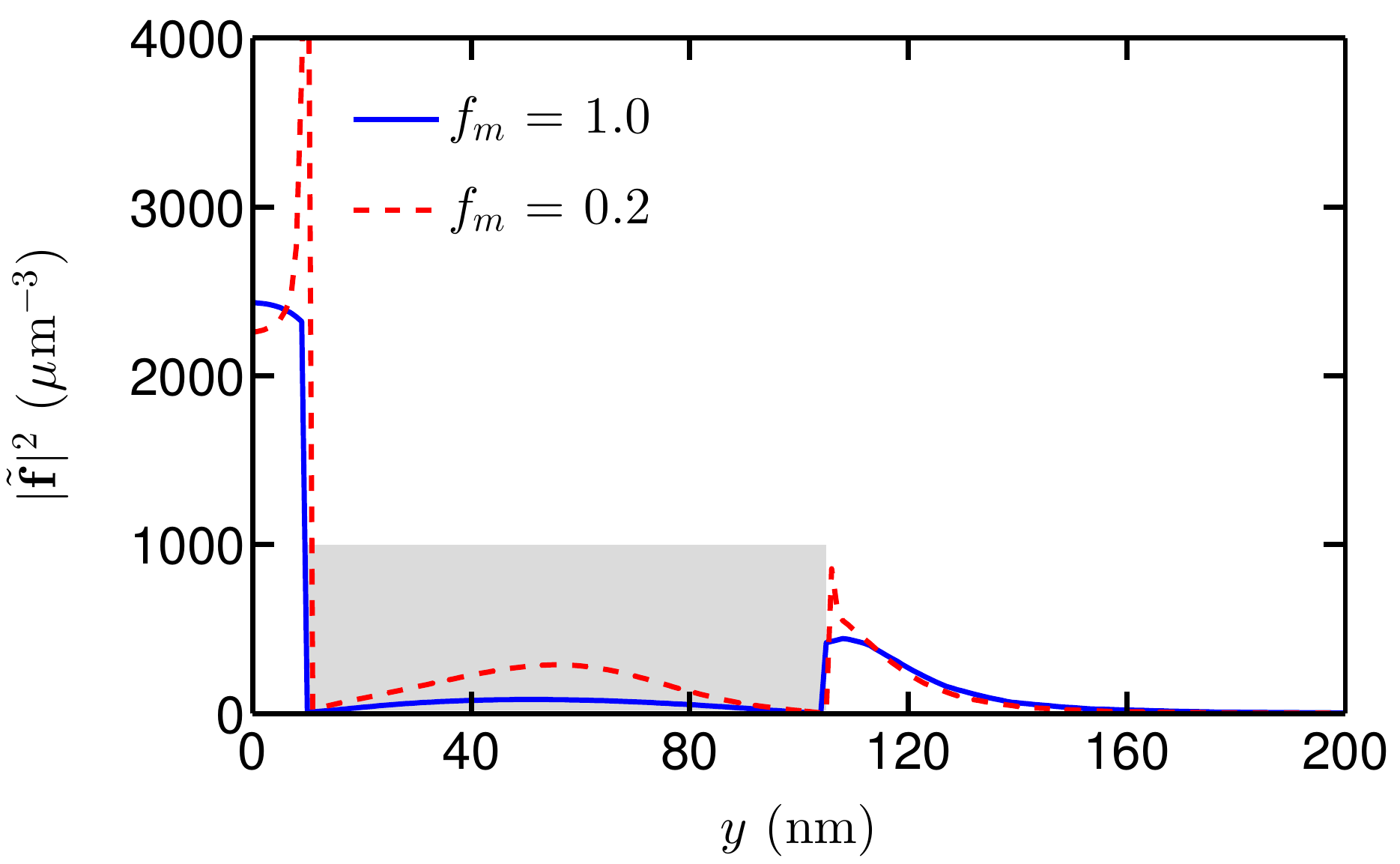}};
        \node (fig2) at (48,24)
        {\includegraphics[width=0.39\columnwidth]{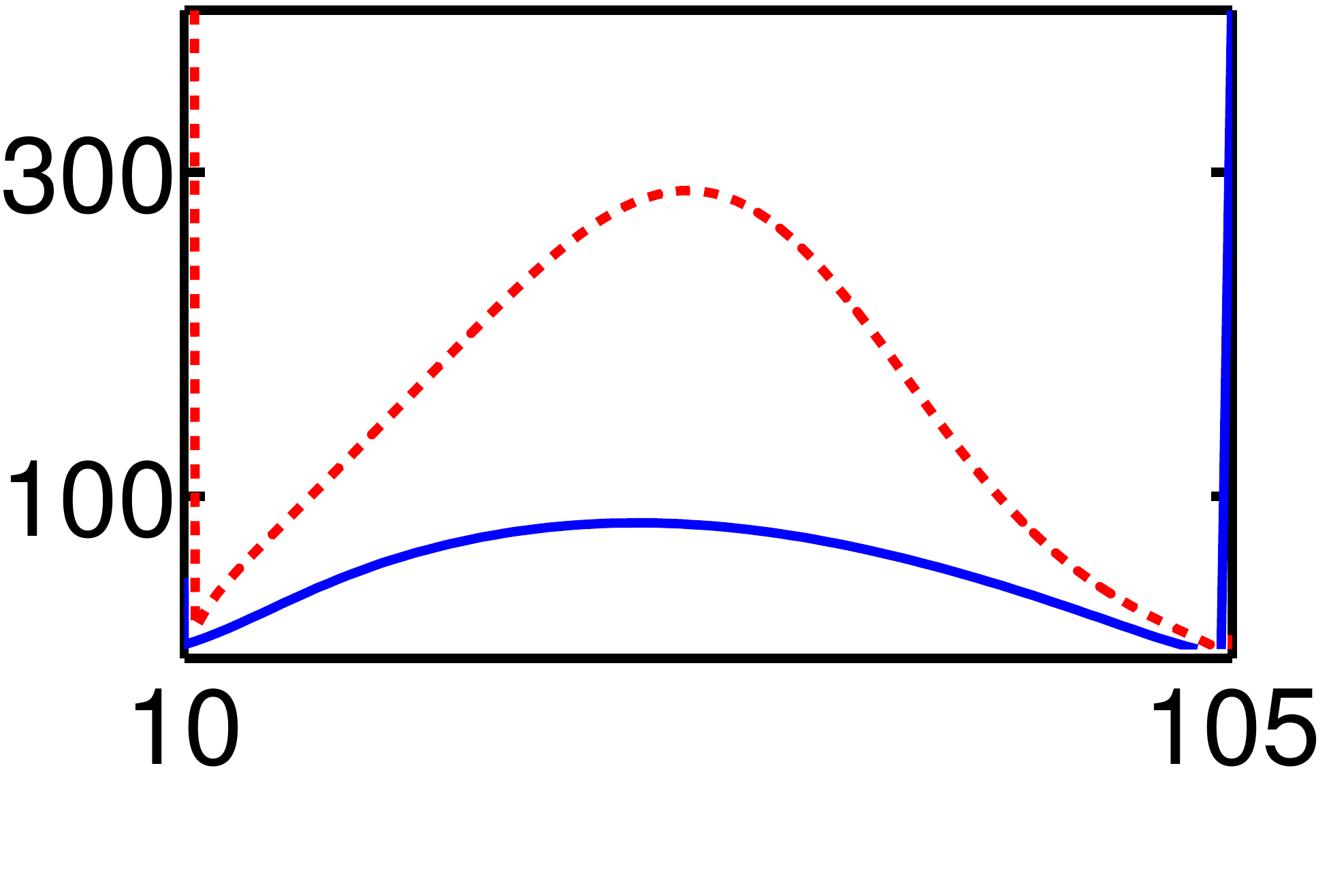}};
        \end{tikzpicture}
          \vspace{-0.6cm}
        \caption{QNM field strength, $\vert \tilde{\mathbf{f}} \vert ^2(0,y,0)$, for the dominant modes of the plasmonic dimer (solid blue) and HMM dimer (dashed red). The shaded region corresponds to positions within each resonator, and specifically to positions within a metal layer of the HMM dimer. Inset: zoom-in of the mode strength for positions inside the resonator.} 
        \label{fig:q_over_v}
\end{figure}
\indent We obtain the QNMs around the resonance of interest for two representative cases: a plasmonic resonator (volume metal filling fraction $f_m = 1.0$), and an HMM resonator with large dielectric character ($f_m= 0.2$). Using COMSOL Multiphysics \cite{comsol}, we use an iterative frequency-domain pole search with a dipole excitation \cite{lalanne_opt_express} to obtain the complex eigenfrequencies, and the associated modes are normalized implicitly. We identify a single complex eigenfrequency for the pure gold dimer, $\tilde{\omega}_c / 2 \pi = 303.29 -i 24.18 \  \mathrm{THz}$ ($Q=6.3$). We obtain a maximum Purcell factor of around $720$ at the origin (gap centre), in excellent agreement with full dipole calculations (Fig. \ref{fig:schematic}c). The HMM dimer response is characterized by three complex eigenfrequencies contributing to the resonance of interest, $\tilde{\omega}_{c_1} / 2 \pi = 139.215 - i 9.847$ THz ($Q=7.1$), $\tilde{\omega}_{c_2} / 2 \pi = 165.335 - i 10.412$ THz ($Q=7.9$), and $\tilde{\omega}_{c_3} / 2 \pi = 197.472 -i 9.860 $ THz ($Q=10.0$). The three-sum QNM maximum Purcell factor at the origin is approximately $5600$, which is within 5\% of the full dipole result of $5900$ (Fig. \ref{fig:schematic}d; the presence of other nearby modes makes the expansion slightly less accurate than that of the gold dimer). 
The HMM and gold QNM profiles are shown in Fig. \ref{fig:qnm_profile}. We remark that the dominant contribution at the origin is from the second QNM, which resembles a plasmonic mode; in contrast, modes 1 and 3 resemble Fabry-Pérot resonances, and contribute strongly at other locations. These results suggest that there is little fundamental difference between plasmonic and HMM modes in nano-resonators, similar to slab structures \cite{2014_prl}.\\
\indent Clearly the Purcell factors achievable with the HMM are much higher than those of the pure gold structure (in this case, by an order of magnitude). However, full dipole calculations yield an impressive $\beta$-factor of  up to $72  \%$ for the metallic resonator, but an extremely poor $\beta$-factor of $12 \%$ for the HMM. We have found similarly low $\beta$-factors for different geometries and configurations, including HMM waveguides, and cylindrical nano-rods and dimers. We have also found low $\beta$-factors in a spherical HMM cavity, in which Ohmic damping was found to decrease with reduced filling fractions \cite{Zhang}, and in an HMM slab structure---see Supplementary Information (SI). To our knowledge, this is the first time that such large losses have been documented in such a wide variety of HMM resonators, and our results stand in contrast to current suggestions in the literature.\\
  \indent \textit{QNM Description of Large Losses.} We argue below that the universally low single photon $\beta$-factors associated with HMMs are attributable to three key factors: (a) HMMs confine light to their metal regions more strongly than metallic resonators, (b) HMM modes have higher quality factors than plasmonic modes, and (c) HMM resonances are redshifted to regimes of higher metallic loss as the metal filling fraction is reduced. In order to understand the first two points, we consider Eq.~(\ref{eq:gamma_nr}) for the case of a $y$-polarized dipole at $\mathbf{r}_a$.  Focusing on a single QNM of interest, the total decay rate is proportional to Im$\{ G_{yy}(\mathbf{r}_a,\mathbf{r}_a; \omega) \} = \mathrm{Im} \{ A(\omega) \tilde{f}_{y}^2(\mathbf{r}_a) \}$, where we have defined $A(\omega) = \omega^2/ 2\tilde{\omega}_{c} (\tilde{\omega}_{c} - \omega)$ for the $c^{\mathrm{th}}$ QNM, and where we have withheld the $c$-dependence of the mode for ease of notation. 
On the other hand, the non-radiative decay rate given by Eq.~(\ref{eq:gamma_nr}) scales with $ \varepsilon'' \vert A(\omega)\vert^2  \vert \tilde{f}_{y}(\mathbf{r}_a) \vert^2  \int_{\mathrm{metal}}   \vert \tilde{\mathbf{f}}(\mathbf{r}) \vert ^2 \mathrm{d} \mathbf{r}$, where $\varepsilon'' = \mathrm{Im} \{ {\varepsilon} \}$, and where we have used vertical bars to indicate both an absolute value and the norm of a vector. For $( \mathrm{Im} \{ \tilde{f} \} / \mathrm{Re} \{ \tilde{f}  \})^2 \ll 1$ and $\mathrm{Im} \{ \tilde{f} \} / \mathrm{Re} \{ \tilde{f}  \} \ll Q$, both of which are almost always satisfied in practice, the on-resonance non-radiative coupling $\eta^{\mathrm{nr}} = \gamma^{\mathrm{nr}} / \gamma $ is given by
\begin{equation}
\eta^{\mathrm{nr}} \propto f_m \varepsilon'' Q \langle \kern 0.13em \vert \tilde{\mathbf{f}} \vert ^2\rangle_{\mathrm{metal}},
\label{eq:gamma_nr_result}
\end{equation}
where $\langle  \kern 0.13em \vert \tilde{\mathbf{f}} \vert^2 \rangle_{\mathrm{metal}} = \int_{V_{\mathrm{metal}}} \vert \tilde{\mathbf{f}}(\mathbf{r}) \vert^2 \ \mathrm{d} \mathbf{r} / V_{\mathrm{metal}}$ denotes an averaging of the field strength over the metal volume, and we have used $V_{\mathrm{metal}} \propto  f_m$ to elucidate the scaling of the non-radiative coupling. Note that the non-radiative decay rate scales with $G^2$, while the total decay rate scales with $G$, so that the non-radiative coupling is increased by an enhancement of $Q \vert \tilde{\mathbf{f}} \vert^2$ within the metal, even if the product increases at $\mathbf{r}_a$ as well. \\
\indent Fig. \ref{fig:q_over_v} shows the mode strength $\vert \tilde{\mathbf{f}} (0,y,0) \vert ^2$ as a function of distance $y$ along the dimer axis, for both $f_m = 1.0$ and $f_m = 0.2$. Outside the dimer, the mode strength is nearly identical for both the metal and the HMM structures, with the only difference occurring a few nm from the metal surface. Within the dimer, however, the mode strength of the HMM is significantly larger than that of the gold resonator. In light of the above discussion, this suggests a much-reduced $\beta$-factor.
\begin{figure}[t]
        \centering
        {\includegraphics[width=0.95\columnwidth]{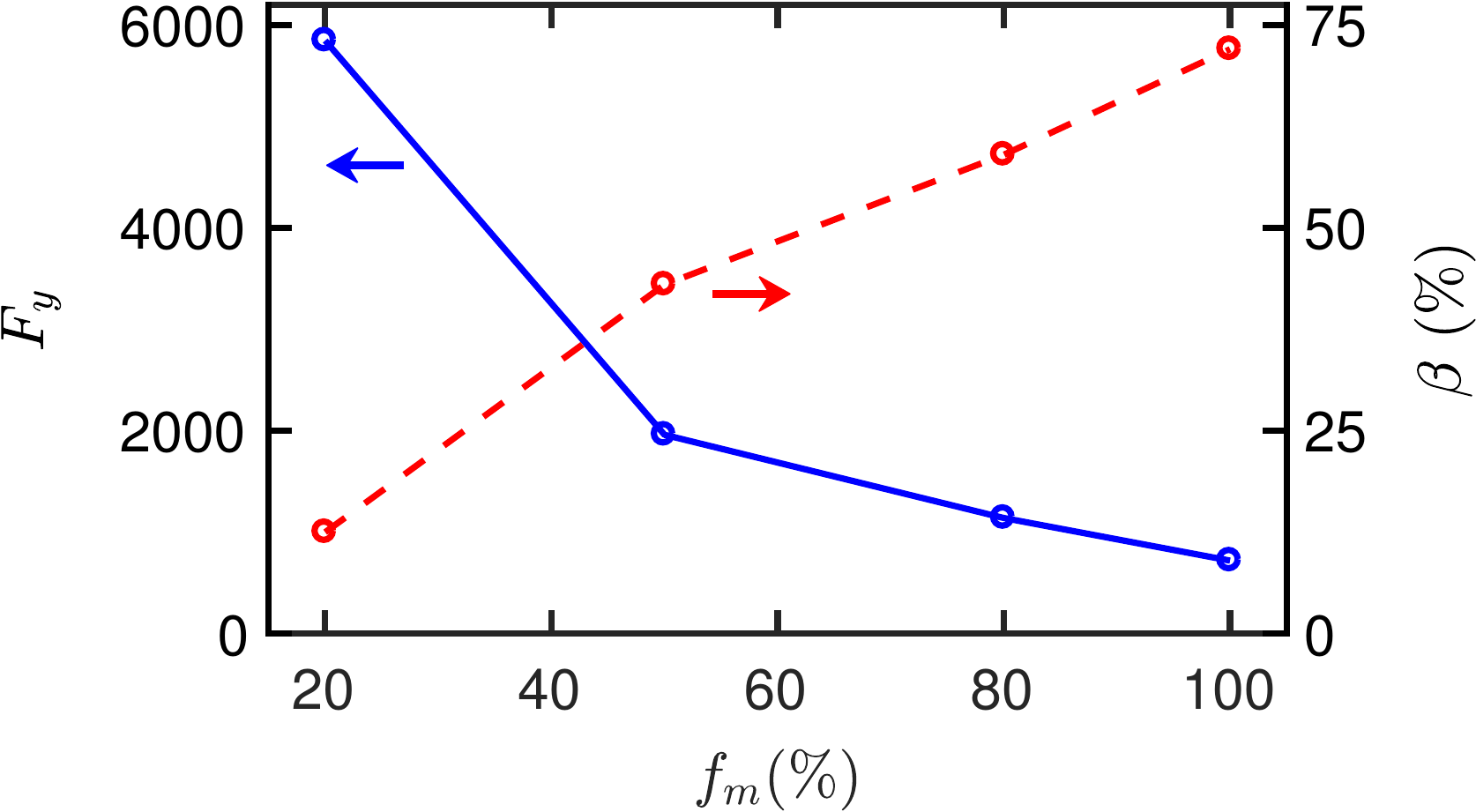}}
          \vspace{-0.0cm}
        \caption{Mid-gap on-resonance Purcell factor (solid blue) and output $\beta$-factor (dashed red) for a $y$-polarized dipole at the origin for varying filling fractions. An increase in the Purcell factor is always accompanied by a reduction in the $\beta$-factor.}
        \label{fig:filling_fraction}
\end{figure}
Evidently, the HMM is not characterized by a stronger modal field at all positions, which would simultaneously increase the Purcell factor while diminishing the $\beta$-factor (see \textit{Comments}  below). In fact, the enhanced light confinement occurs only within the structure. We can understand this effect as arising from the increased quantity of dielectric within the resonator. Since the dielectric supports the existence of electric fields better than the metal, the field strength within the structure becomes stronger as the metal volume is reduced. The field strength is enhanced in both dielectric and metal layers, and the latter effect leads to increased loss (see SI for more details). Such an explanation suggests that smaller metal filling fractions are associated with higher loss, which is indeed observed (Fig. \ref{fig:filling_fraction}). We suggest that this effect is characteristic of all resonators consisting of metal and dielectric layers, and it is indeed consistent with all cases we have studied.\\
\indent The $\beta$-factor is further reduced by an increase in quality factor, $Q^{\mathrm{HMM}} / Q^{\mathrm{metal}} = 1.27$, and from the redshifting of the resonance frequency, since metals with a Drude-like dispersion are characterized by a loss term $\varepsilon'' \propto 1/\omega^3$. Importantly, this latter effect balances the reduction in the metal volume, such that the product $f_m \varepsilon''$ appearing in Eq. \ref{eq:gamma_nr_result} is equal to 3.16 for the HMM dimer, and to $2.59$ for the gold dimer (see SI). This balancing effect, combined with increased $Q$-factors and enhanced light confinement within the metal, leads to lower $\beta$-factors associated with the HMM dimer. \\  
\indent In light of the above results, we suggest that HMM resonators make poor single photon sources, for any Purcell factor improvement over metal resonators is accompanied by a reduction in the $\beta$-factor (which renders the photon source increasingly non-deterministic). This surprising result is expected to be true of all forms of HMM nano-resonators, given the general form of the explanation given above, and we have found it to be true in all of the examples we have studied. Our results thus suggest that HMM structures may be limited by non-radiative loss in ways that pure metal structures are not.  \\
\indent \textit{Comments.} As seen in Fig. \ref{fig:q_over_v}, the QNM strength of the HMM resonator is no larger than that of the metal. The increased $Q$-factor of the HMM yields a small enhancement in the GF, but the effect is rather minor. In fact, the superior HMM Purcell factor is largely due to a decreased resonance frequency. Since the free-space decay rate of a dipole emitter scales with $\omega^3$ \cite{novotny}, the associated spontaneous emission enhancement is larger at lower frequencies. Evidently HMMs cannot access non-perturbative quantum optics effects such as the strong coupling regime and vacuum Rabi splitting, which rely on an enhanced GF \cite{novotny}, unless they can also be accessed by metals. Indeed, we have found that vacuum Rabi splitting for a typical quantum dot dipole requires Purcell factors that are orders of magnitude larger than any of the enhancements found here. These results are consistent with those obtained for HMM slab structures \cite{2014_prl}. While the strong resonance redshift associated with decreased filling fractions provides an opportunity to finely tune to dipole resonances, such tuning may also be possible by modifying the size of metal resonators \cite{2014_prl}.\\
\indent \textit{Conclusions.} We have shown that coupling to HMM nano-resonators can lead to Purcell enhancements that are much larger than those of metals with comparable geometries. Surprisingly, however, we have found that these enhancements are associated with unusually low $\beta$-factors. Using a semi-analytical QNM approach, we have shown that these low $\beta$-factors are due to redshifted resonances, increased quality factors, and stronger confinement of light within the metal. We conclude that HMM nano-resonators are poor choices for single photon sources and other applications requiring strong radiative coupling, though they undoubtedly have other uses and advantages in
 other areas of plasmonic quantum  optics.

This work was supported by the Natural Sciences and
Engineering Research Council of Canada. We acknowledge CMC Microsystems for the provision of COMSOL Multiphysics to facilitate this research, and  thank
Rongchun Ge for useful discussions.

\end{document}


\captionsetup[subfigure]{labelformat=empty}
        \frenchspacing 
        \abovedisplayskip=8pt
        \abovedisplayshortskip=8pt
        \belowdisplayskip=8pt
        \belowdisplayshortskip=8pt
        \arraycolsep=100pt
        
      \title{Hyperbolic Metamaterial Nano-Resonators Make Poor Single Photon Sources: Supplementary Information}
      
      \author{Simon Axelrod}
      \affiliation{Department of Physics, Engineering Physics and Astronomy,
        Queen's University, Kingston, Canada, K7L 3N6}
      
      \author{Mohsen Kamandar Dezfouli}
      \affiliation{Department of Physics, Engineering Physics and Astronomy,
        Queen's University, Kingston, Canada, K7L 3N6}
      
      \author{Herman M. K. Wong}
      \affiliation{Photonics Group, Edward S. Rogers Sr. Department of Electrical and Computer Engineering, University of Toronto, Toronto, Canada, M5S 3H6}
      
      \author{Amr S. Helmy}
      \affiliation{Photonics Group, Edward S. Rogers Sr. Department of Electrical and Computer Engineering, University of Toronto, Toronto, Canada, M5S 3H6}
      
      \author{Stephen Hughes}        
      \email{shughes@queensu.ca}
      \affiliation{Department of Physics, Engineering Physics and Astronomy,
        Queen's University, Kingston, Canada, K7L 3N6}

      \begin{abstract} \noindent Here we provide supplementary material that accompanies the manuscript ``Hyperbolic Metamaterial Nano-Resonators Make Poor Single Photon Sources.'' Specifically, we compare the Purcell and 
single photon $\beta$-factors for an HMM and metal slab structure, as well as an HMM sphere supporting whispering gallery resonances. We also supply the main simulation parameters and software tools used in our numerical simulations.
 We demonstrate that the behavior of the Purcell and $\beta$-factors is the same as for the resonators described in the main text. We also motivate our results regarding the resonance frequency scaling in the main text, using a quasi-static model of an HMM spherical resonator. We argue that the resonance condition implies that the product $f_m \varepsilon''$ should increase as the metal filling fraction is reduced. Next, we show that the low $\beta$-factors of HMM nano-resonators are consistent with the prediction of a simple quasi-static analysis. Finally, we analyze the multi-mode behaviour of the HMM nano-resonator of the main text.       \end{abstract}
      
      \maketitle
      
\renewcommand{\thefigure}{S\arabic{figure}}
\renewcommand{\theequation}{S\arabic{equation}}

\begin{figure}[b]
        \centering
        \begin{tikzpicture}
        \node[anchor=south west,inner sep=0] at (-5.8,0.5){\includegraphics[width=0.2692\columnwidth,trim={0cm 0cm 0cm 0cm},clip]{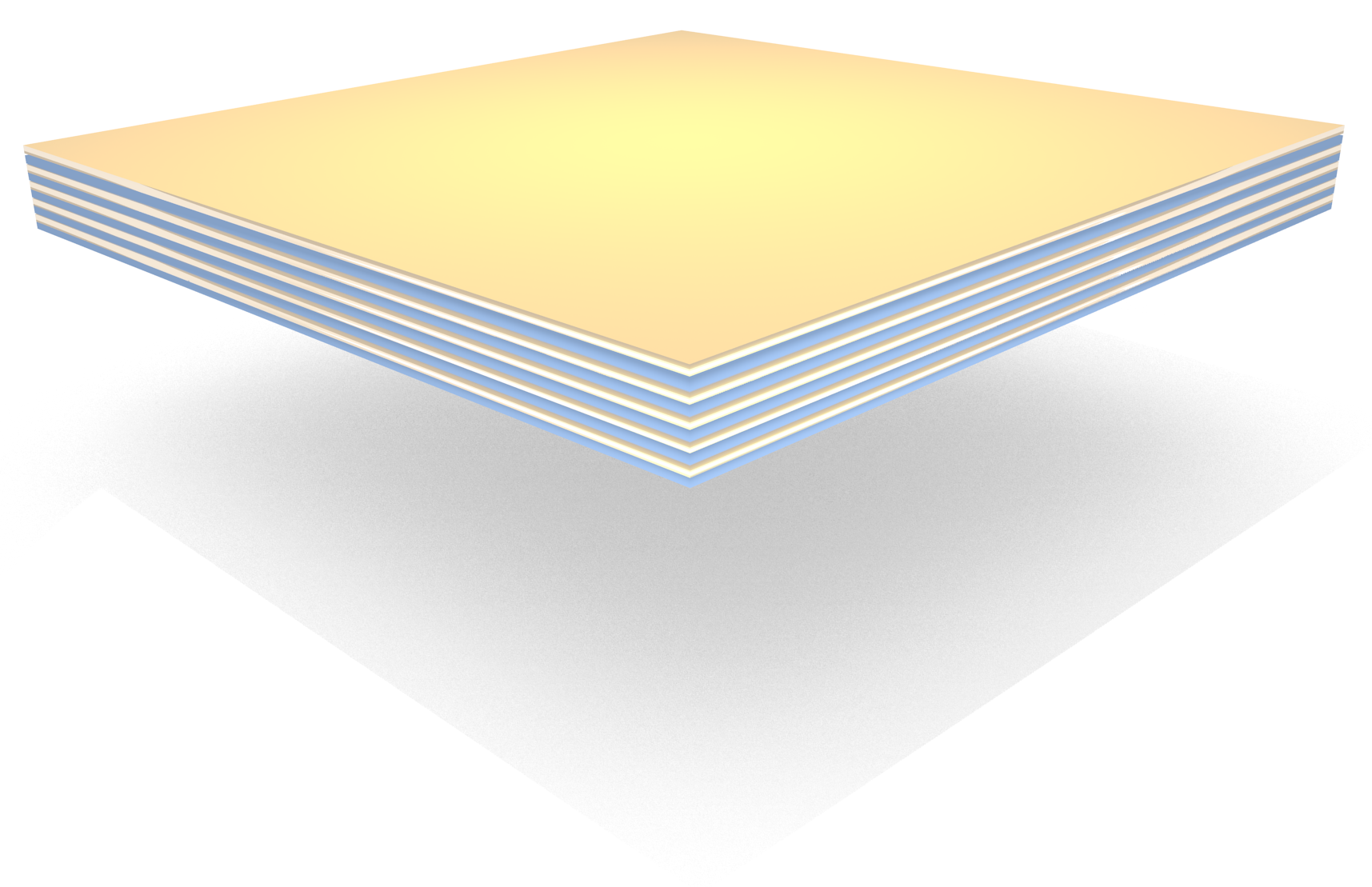}};
        \node[anchor=south west,inner sep=0] at (5.8,0.2){\includegraphics[width=0.32\columnwidth,trim={0cm 0cm 0cm 0cm},clip]{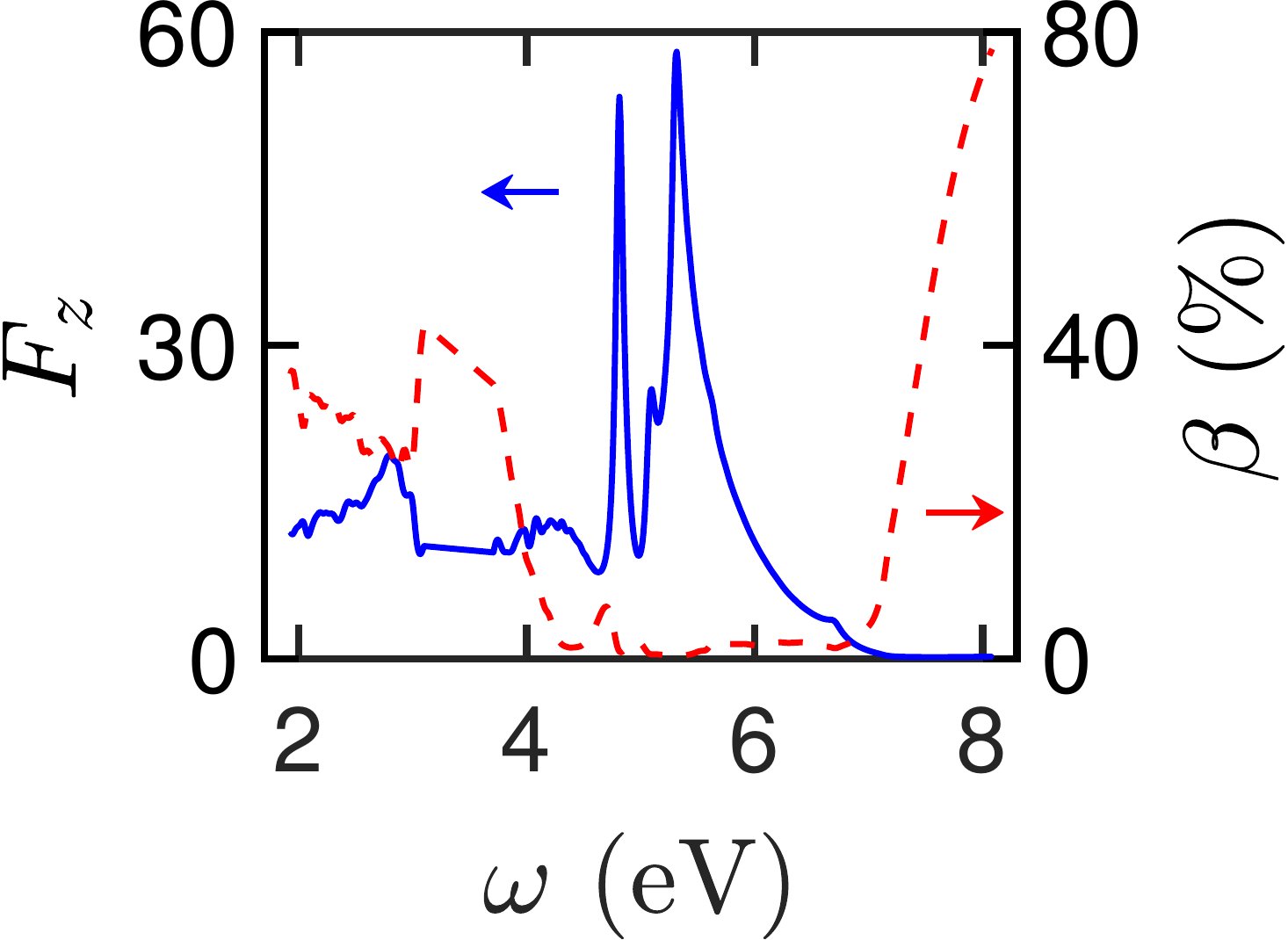}};
     
        \node[anchor=south east,inner sep=0] at (5.5,0.2){\includegraphics[width=0.32\columnwidth,trim={0cm 0cm 0cm 0cm},clip]{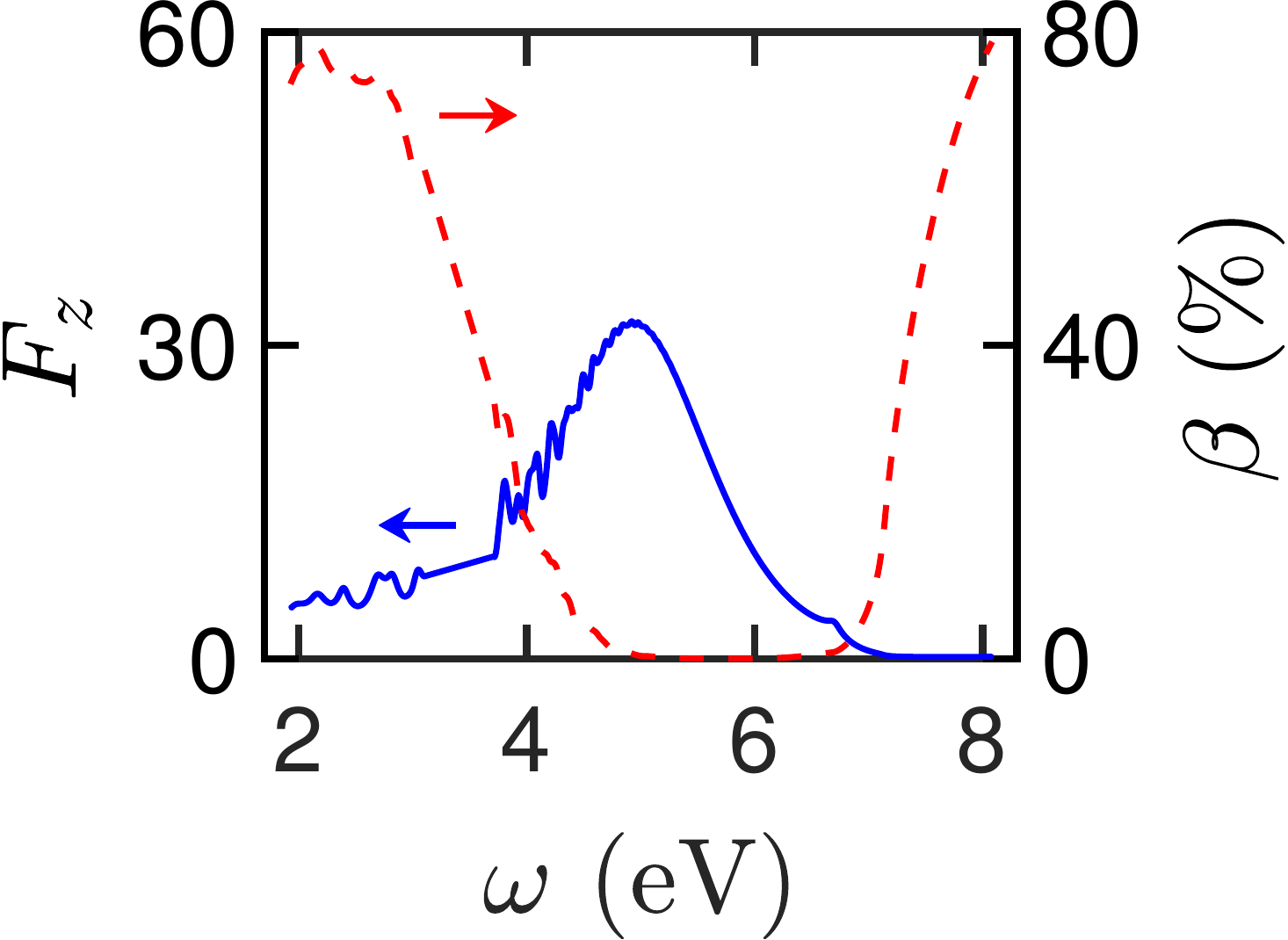}};
        
        \node at (-6,3.75) {(a)};
        \node at (0.3,3.75) {(b)};
                \node at (6.2,3.75) {(c)};
        \end{tikzpicture}
        \caption{Comparison of single photon parameters of a gold slab and an HMM slab of 50\% metal filling fraction. (a) Schematic of an HMM slab. (b) Purcell factor and $\beta$-factor for a vertically-polarized dipole located 10 nm from the gold slab surface. (c) As in (b), but for an HMM slab of 50\% metal filling fraction. For both structures, we see a clear correspondence between Purcell and $\beta$-factors. In particular, the $\beta$-factors are vanishingly small near the main resonances.}
        \label{fig:beta}
\end{figure}

\begin{figure}[t]
        \centering
        \begin{tikzpicture}
        \node[anchor=south east,inner sep=0] at (10,0){\includegraphics[width=0.43\columnwidth,trim={1cm 0cm 0cm 1cm},clip]{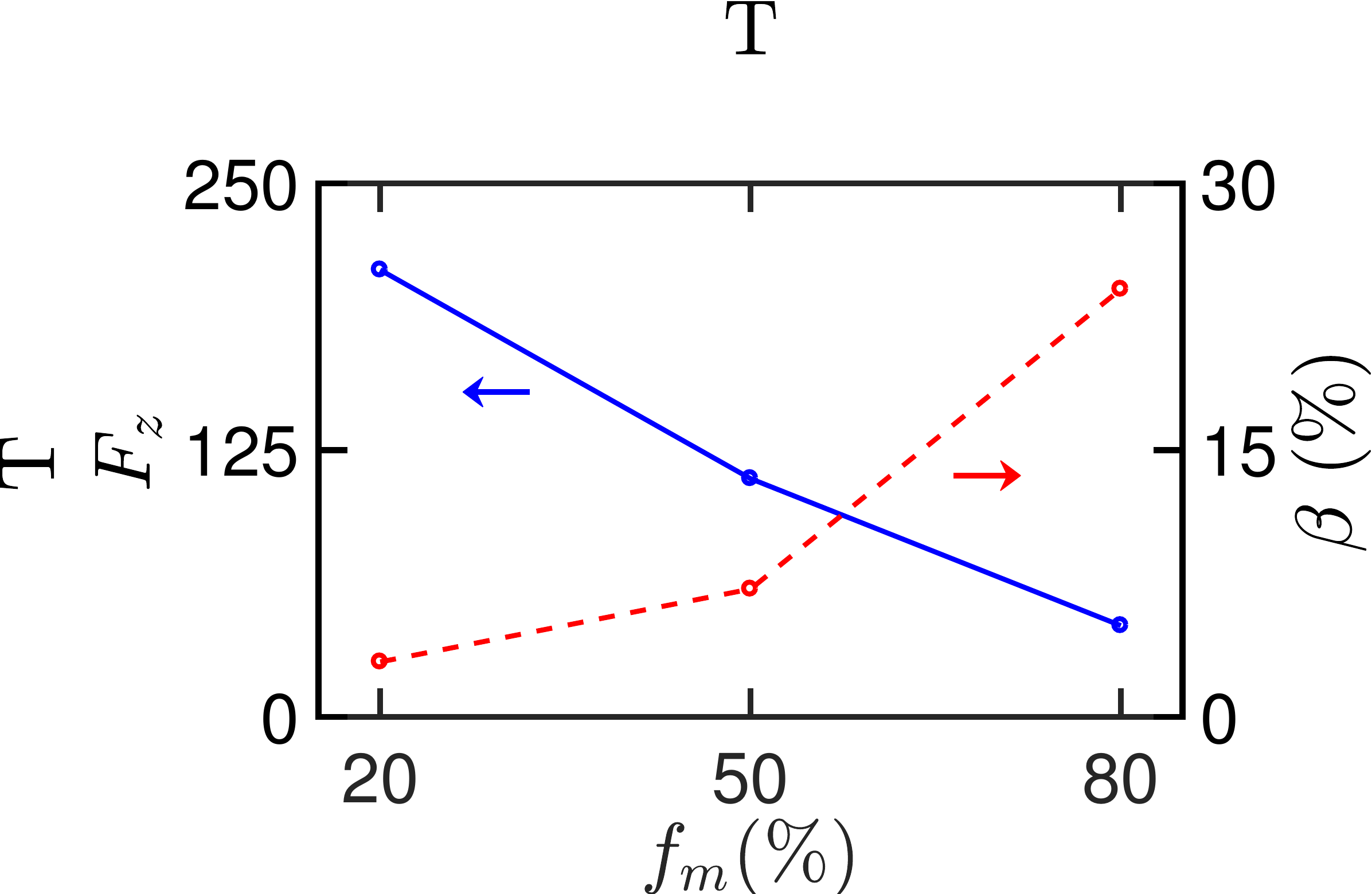}};
        \node[anchor=south west,inner sep=0] at (-3.6,0.1){\includegraphics[width=0.27\columnwidth,trim={0cm 0cm 0cm 0cm},clip]{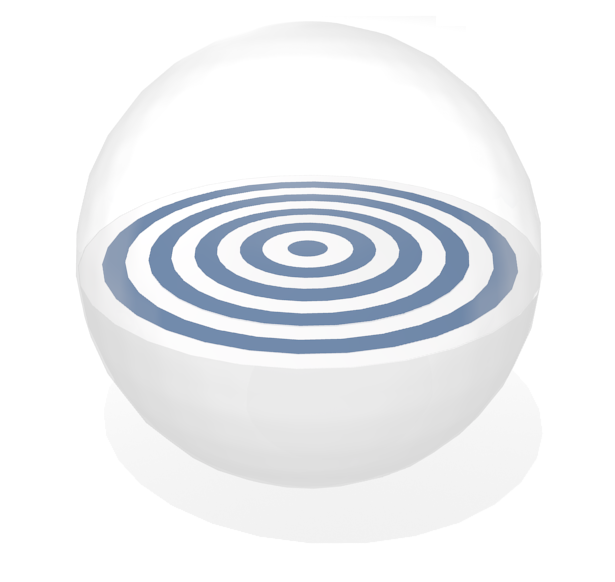}};
        \node at (-3.7,3.4) {(a)};
        \node at (2,3.4) {(b)};
        \end{tikzpicture}
         \caption{(a) Schematic of an HMM micro-sphere. (b) Purcell factor (solid blue) and $\beta$-factor (dashed red) as a function of metal filling fraction for a $z$-polarized dipole located 10 nm from the outer surface of the sphere. For the larger Purcell factors achievable inside the HMM dielectric regions (not shown), the single photon  $\beta$-factors are negligible.}
        \label{fig:zhang}
\end{figure}
        
        
\section{Purcell and $\beta$-Factors for Two Different HMM Structures}
To provide additional generality to the results in the main text, here we present computational results for the Purcell and $\beta$-factors associated with two completely different HMM and metal structures, and show that they are consistent with those of the nano-resonators studied in the main text. We use a slab structure characterized by a continuum of modes to show that low $\beta$-factors seem to be a general feature of HMM structures, and are not limited to HMM nano-resonators in particular (though we note that the multi-mode behaviour of slab structures is problematic for single photon applications). We also examine the spherical HMM resonator studied in Ref. \onlinecite{Zhang} to further support the argument that HMM nano-resonators make poor single photon sources. \\ \indent First we compare a gold slab with an HMM slab of 50\% metal filling fraction. The width and length of the slab are one micron ($x$- and $y$-directions), and its height is 150 nm ($z$-direction). The HMM consists of 5 layers of gold and 5 layers of dielectric, each with a thickness of 15 nm, and we use the same parameters for the dielectric constants as in the main text. We calculate the $\beta$-factors and Purcell factors as a function of frequency for a $z$-polarized dipole located 10 nm from the surface, through a full dipole calculation 
using Lumerical finite-difference time-domain (FDTD) simulations~\cite{lum1,lum2}.
Since we scan a large region of frequency space without an obvious modal structure it is more convenient to use FDTD for these calculations. 
The FDTD simulations  were performed using a 5 nm mesh within a 2 $\mu {\rm m}^3$ computational domain, excluding the 64 perfectly matched layers (PMLs) used to simulate the outgoing boundary condition.
The results are shown in Fig. \ref{fig:beta}. Well below the plasma frequency, the Purcell factor of the HMM is about double that of the gold slab, while the $\beta$-factor of the gold slab is much higher than that of the HMM (around 80\% versus 40\%). There are higher frequencies which the HMM Purcell factors are larger, and others for which the gold Purcell factors are larger. However, it is important to note that in these ranges the $\beta$-factors of each are vanishingly small. In all cases, any enhancement in the Purcell factor is associated with a decrease in the $\beta$-factor, which is in agreement with the conclusion made in the main text.
 \\
\indent It is also important to note that the Purcell factors obtained here are orders of magnitude smaller than those of nano-resonator structures. Moreover, it is clear that the Purcell factors represent contributions from a number of resonant modes. A typical requirement for an ideal single photon source is that dipole emitters couple to a single mode only, with $\beta$- and Purcell factors that are as large as possible. It would thus be preferable, and likely necessary, to use nano-resonators in place of slab structures for such applications. In this context, it is highly desirable to have a modal picture of the underlying physics, in much the same way that one typically analyzes microcavity-enabled cavity-QED effects. \\ 
\indent Next we investigate the HMM micro-sphere studied in Ref. \onlinecite{Zhang}, which supports whispering gallery resonances. For this structure, we have used COMSOL Multiphysics, as in the main text \cite{comsol}.
The COMSOL calculations for both the cylindrical resonators (studied in the main text) and spherical geometries (shown here) were performed within a 0.2 $\mu {\rm m}^3$  computational domain for all filling fractions. This domain size included all PML layers. The number of computational elements used for each structure was different in order to meet the different geometrical demands. A minimum of 70,000 elements were used for simulations of pure gold structures, while a maximum of 200,000 elements were used for low filling fraction HMMs. In addition, 10 layers of PML were used in all calculations, which were enough to obtain accurate numerical convergence. 
 The HMM\ sphere has a radius of 100 nm, and consists of 5 layers of silver and 5 layers of dielectric; further details can be found in Ref. \onlinecite{Zhang}. We obtain $\beta$-factors and Purcell factors for a $z$-polarized dipole located at $z$=$10$ nm from the surface of the sphere, coupling to the angular momentum $l=2$ mode. The results shown in Fig. \ref{fig:zhang} mirror those of the resonator studied in the main text: the Purcell factor increases and the $\beta$-factor decreases as the filling fraction is reduced. These results are consistent with our general conclusions about non-radiative decay in HMM resonators. As well, it was concluded in Ref. \onlinecite{Zhang} that Ohmic damping decreases as the filling fraction is reduced, leading to increased quality factors. Evidently this does not lead to less Ohmic loss, for the $\beta$-factor is reduced for smaller filling fractions. This result is consistent with Eq.~(3) in the main text, which shows that $\eta^{\mathrm{nr}}$ is actually proportional to $Q$.\\

\section{Resonance Frequency Scaling}
In the main text we argue that the enhanced Purcell factors in HMM nano-resonators are mainly due to a resonance frequency redshift. We note that this redshift leads to a larger loss term through the enhancement of the imaginary part of the dielectric constant. This enhancement is such that that the product $f_m \varepsilon''$ appearing in Eq.~(3) of the main text is actually increased. Here we further motivate this result with a simple example. \\
\indent One can analyze a spherical HMM nano-resonator in the quasi-static approximation, using an effective medium description (see Ref. \onlinecite{Zhang} for the form of the model used). For a Drude metal and dielectric layers with unit permittivity, the resonance condition is found to be $\omega_0 = \omega_p \sqrt{f_m/3}$. Clearly the resonance frequency is a decreasing function of metal filling fraction. Moreover, an application of the Drude formula shows that the product $f_m \varepsilon''$ increases as the filling fraction is reduced. This is a direct result of the fact that the imaginary part of $\varepsilon$ scales $1/\omega^3$, while the real part scales as $1/\omega^2$. This implies that the redshift accompanying the increased Purcell factor yields an increased loss parameter that is large enough to balance the decrease in filling fraction. More generally, one expects that a plasmonic resonance will occur when a denominator of the form $\varepsilon + \alpha \epsilon^h$ becomes resonant, for some $\alpha$ that depends on the given configuration. For an HMM described as an effective medium, the metal component of the permittivity is given as $\varepsilon = f_m \varepsilon_m + (1-f_m) \varepsilon_d$ \cite{Zhang}. Satisfying the resonance condition then implies that $\omega_0$ is a decreasing function of $f_m$, and an application of the Drude formula shows that the product  $f_m \varepsilon''$ must increase as the filling fraction is reduced. \\

\section{Quasi-Static Picture of Diminishing HMM $\beta$-Factors}
We follow the approach taken in Ref. \onlinecite{qs_energy}, which makes use of a quasi-static approximation, deemed to be valid for resonators whose dimensions are much smaller than the resonant wavelength. Such an approach becomes increasingly well-justified for HMM nano-resonators, as the size of the resonator remains constant while the resonance frequency is reduced. In the quasi-static limit, the localized modes of a resonator are bound by the following relation: 
\begin{align}
\int_{V_m} -\varepsilon_m' \vert \tilde{\mathbf{F}} (\mathbf{r}) \vert ^2 \  \mathrm{d} \mathbf{r}=  \int_{V_d} \varepsilon_d(\mathbf{r}) \vert \tilde{\mathbf{F}} (\mathbf{r}) \vert ^2 \  \mathrm{d} \mathbf{r}.
\label{equiv}
\end{align}
Here,  $\tilde{\mathbf{F}}(\mathbf{r})$ is a ``localized field mode", $\varepsilon' = \mathrm{Re} \{\varepsilon\}$, $V_m$ is the metal volume, and $V_d$ is the total dielectric volume (including the volume of the dielectric component of the resonator). The localized mode is defined here as \cite{njp} 
\begin{align} 
\tilde{\mathbf{F}}(\mathbf{r}) = \int_{V_{\mathrm{dimer}}} \mathbf{G}^h\mathbf{(r,r';\omega)} \cdot \Delta \varepsilon(\mathbf{r}', \omega) \ \tilde{\mathbf{f}}(\mathbf{r}') \ \mathrm{d} \mathbf{r}', \ \ \ &\mathrm{outside \ the  \ dimer} \nonumber \\
= \tilde{\mathbf{f}}(\mathbf{r}), \ \ \ &\mathrm{inside \ the  \ dimer}.
\end{align} Here, $\mathbf{G}^h\mathbf{(r,r';\omega)}$ is the Green function of the homogeneous background medium, $\Delta \varepsilon(\mathbf{r}')$ is the permittivity shift within the dimer, and $\tilde{\mathbf{f}}{(\mathbf{r}')}$ is the QNM (see main text). This localized field mode is essentially a regularized QNM, which corresponds to the QNM at positions near the resonator, but does not diverge in the far field~\cite{njp}.
 \\  \indent Invoking the Drude formula for $\omega \ll \omega_p$, and using Eq. (2) of the main text, we obtain the non-radiative decay rate for an $\mathbf{n}_a$-polarized dipole emitter at position $\mathbf{r}_a$:
\begin{align}
\label{eq:gamma_nr}
& \gamma^{\rm{nr}}(\mathbf{r}_{a},\omega) = \frac{2d^2 \gamma_{\mathrm{col}} \vert A(\omega) \tilde{F}_a(\mathbf{r}_a) \vert ^2}{\hbar \omega \varepsilon_0} \int_{V_d} \varepsilon_d(\mathbf{r}) \vert \tilde{\mathbf{F}} (\mathbf{r}) \vert ^2 \  \mathrm{d} \mathbf{r},
\end{align}
where  $\gamma_{\mathrm{col}}$ is the collision damping rate in the Drude formula. The on-resonance $\beta$-factor is then
\begin{align}
\label{beta}
\beta= 1- Q \frac{\gamma_{\mathrm{col}}}{\omega} \int_{V_d} \varepsilon_d(\mathbf{r}) \vert \tilde{\mathbf{F}} (\mathbf{r}) \vert ^2 \  \mathrm{d} \mathbf{r}.
\end{align}
We see that the $\beta$-factor decreases as the integrated mode strength over the total \textit{dielectric} volume increases, and as the resonance frequency is reduced. This is precisely what we have observed in HMMs: as the dielectric volume increases, and the resonance frequency drops, the $\beta$-factor decreases. The physical justification for this effect is the same as the one given in the main text. The $\omega^{-1}$ pre-factor reflects the fact that lower frequency regimes are associated with larger loss,
while the integral of the field strength over the dielectric regions reflects the fact that stronger fields in the dielectric lead to stronger fields in the metal, and thus to larger losses, as well.
\\ \indent Note that this behaviour is different from that of a plasmonic resonator of reduced volume. As the volume is reduced in an ordinary resonator, the resonance frequency becomes blue-shifted. However, the smaller volume of the resonator leads to enhanced field strengths both inside and outside the resonator, and thus to larger loss. Both HMM and metal resonators are limited in their increased Purcell enhancement by a reduction in the $\beta$-factor, but the reasons for each are subtly different. \\
\begin{figure}[h]
        \centering
          \centering
          \begin{tikzpicture}
          \node[anchor=south west,inner sep=0] at (5.8,0.207){\includegraphics[width=0.3725\columnwidth,trim={0cm 0cm 0cm 0cm},clip]{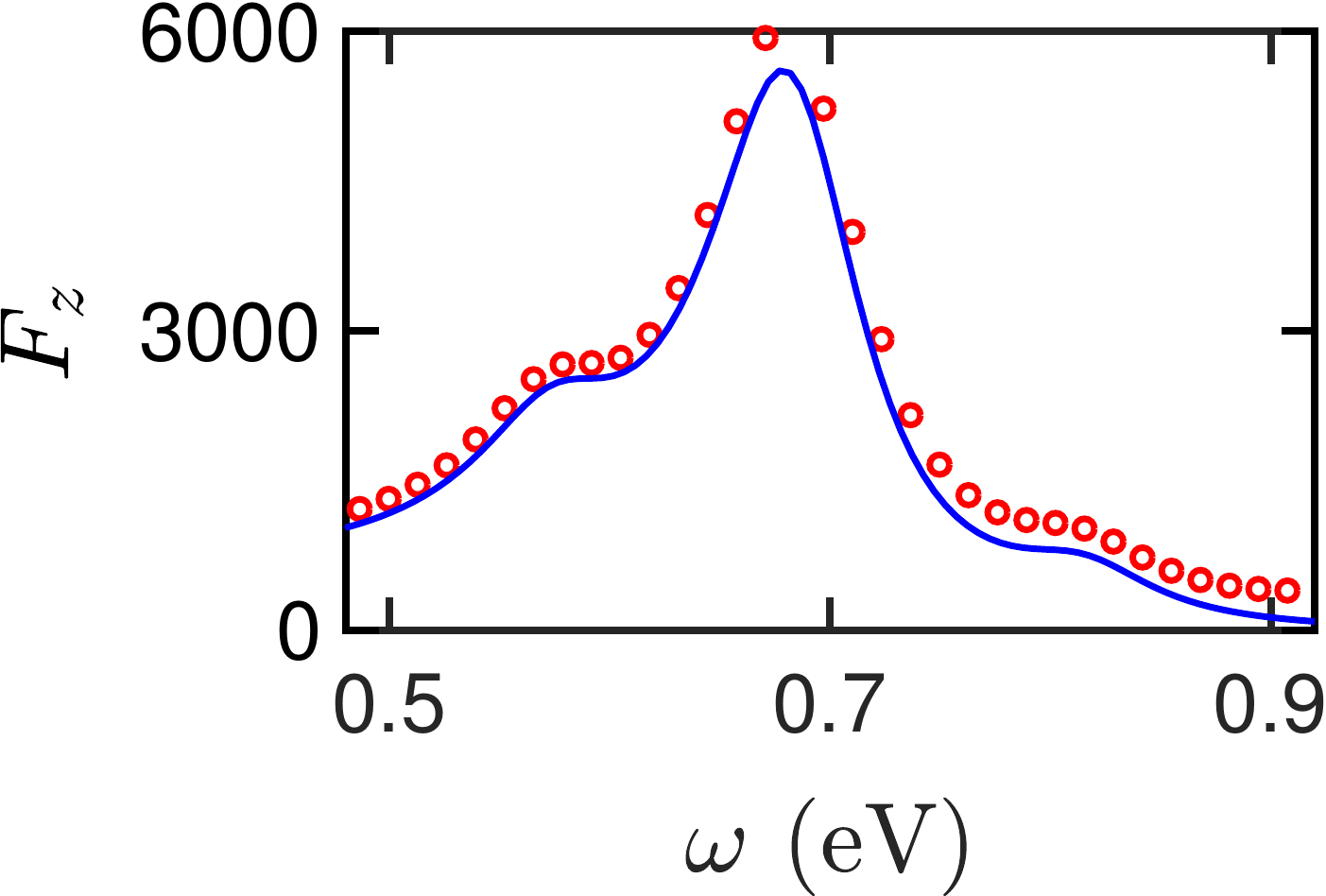}};
          
          \node[anchor=south east,inner sep=0] at (4,0.2){\includegraphics[width=0.37\columnwidth,trim={0cm 0cm 0cm 0cm},clip]{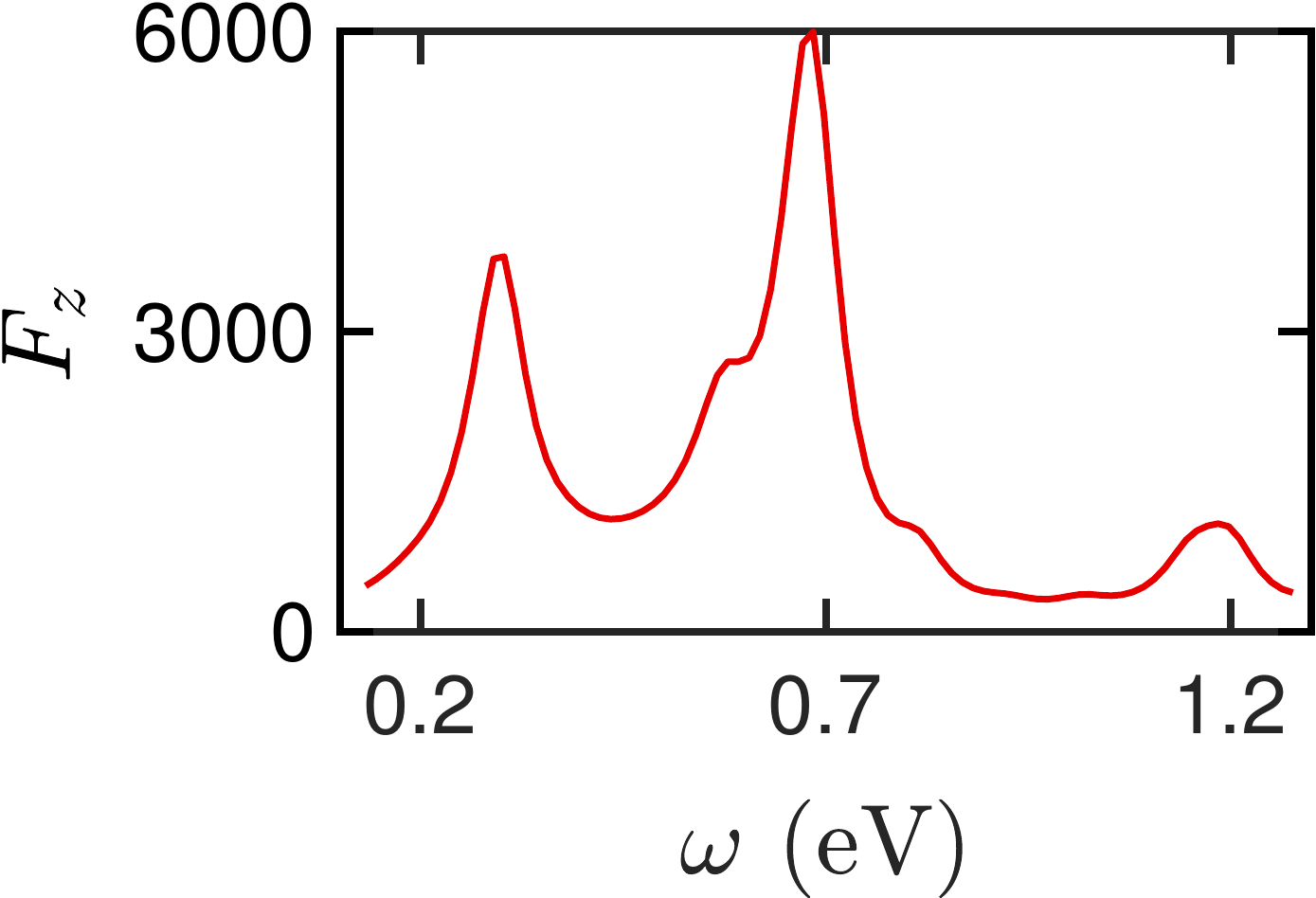}};
           \node at (5,3.75) {(b)};
           \node at (-3.5,3.75) {(a)};
         \end{tikzpicture}
        \caption{(a) Extended view of the Purcell factor associated with the HMM nano-resonator studied in the main text, as calculated through full dipole simulations. (b) Purcell factor in the resonant regime of interest, as calculated through full dipole simulatons (red circles) and an expansion of three QNMs (solid blue).  }
        \label{fig:multimode}
\end{figure}
\section{Multi-Mode Behaviour of HMM Nano-Resonators} In the main text we note that the presence of nearby modes makes the QNM expansion slightly less accurate for the HMM resonator, which may seem surprising given the excellent accuracy of the plasmonic QNM result. For completeness we have included an extended view of the HMM Purcell factor in Fig. \ref{fig:multimode}, as calculated through full dipole simulations. It is clear that, in addition to the main plasmonic peak near 0.7 eV, as well as the accompanying Fabry-P\'erot resonances, there are also interfering modes at higher and lower frequencies. Nevertheless, the three QNM expansion used in the main text is accurate to within 5\% in the region of interest near the main peak, as seen in Fig.  \ref{fig:multimode}b. 